\documentclass[%
 reprint,
 amsmath,amssymb,
 aps,
]{revtex4-1}

\usepackage{graphicx}% Include figure files
\usepackage{dcolumn}% Align table columns on decimal point
\usepackage{bm}% bold math
\usepackage{textcomp}
\begin{document}

%\preprint{APS/123-QED}

\title{Transition between Kerr comb and stimulated Raman comb in a silica whispering gallery mode microcavity}% Force line breaks with \\

\author{Shun Fujii}
\affiliation{Department of Electronics and Electrical Engineering, Faculty of Science and Technology, Keio University, Yokohama, 223-8522, Japan}

\author{Takumi Kato}
\affiliation{Department of Electronics and Electrical Engineering, Faculty of Science and Technology, Keio University, Yokohama, 223-8522, Japan}

\author{Ryo Suzuki}
\affiliation{Department of Electronics and Electrical Engineering, Faculty of Science and Technology, Keio University, Yokohama, 223-8522, Japan}

\author{Atsuhiro Hori}
\affiliation{Department of Electronics and Electrical Engineering, Faculty of Science and Technology, Keio University, Yokohama, 223-8522, Japan}%

\author{Takasumi Tanabe}
\email{takasumi@elec.keio.ac.jp}
\affiliation{Department of Electronics and Electrical Engineering, Faculty of Science and Technology, Keio University, Yokohama, 223-8522, Japan}


%\affiliation{
%\affiliation{
%\affiliation{%

\date{\today}% It is always \today, today,

\begin{abstract}
We theoretically and experimentally investigated the transition between modulation instability and Raman gain in a small silica microcavity with a large free-spectral range (FSR), which reveals that we can selectively switch from a four-wave mixing dominant state to a stimulated Raman scattering dominant state. Both the theoretical analysis and the experiment show that a Raman-dominant region is present between transitions of Kerr combs with different free-spectral range spacings. We can obtain a stable Kerr comb and a stable Raman state selectively by changing the driving power, coupling between the cavity and the waveguide, and laser detuning. Such a controllable transition is achieved thanks to the presence of gain competition between modulation instability and Raman gain in silica whispering gallery mode microcavities.
%\begin{description}
%\end{description}
\end{abstract}

\maketitle

%\tableofcontents

\section{Introduction}

Intracavity nonlinear frequency conversions have been intensively investigated using whispering gallery mode (WGM) microcavities and microring resonators. A high-quality factor ($Q$) and a small mode volume ($V$) allow us to obtain optical nonlinearities at a very low input power because the electrical field is greatly enhanced inside the cavity. Third-order nonlinearities such as four-wave mixing (FWM), third-harmonic generation (THG), stimulated Raman scattering (SRS), and stimulated Brillouin scattering (SBS) are observed and studied. In particular, cascaded FWM is known as a basic mechanism for the generation of a Kerr comb~\cite{DelHaye2007,kippenberg2011microresonator,PhysRevLett.93.083904,PhysRevLett.101.093902,Ferdous2011,Levy2010}, which is expected to be used for various applications such as microwave generation~\cite{Liang2015:high}, spectroscopy~\cite{Suh600}, and optical clocks~\cite{papp2014microresonator}. A better understanding of Kerr comb generation obtained theoretically and experimentally will boost the use of this technology.

In addition to Kerr comb generation, much attention has been paid to SRS in microcavity systems made of different materials~\cite{Grudinin:07,Latawiec:15,Griffith:16,Lin:16phase}. SRS is explained as an interaction between a pump photon $\omega_p$ and a Stokes photon $\omega_s$ where the frequency difference $\omega_p-\omega_s$ matches the molecular vibration frequency $\omega_v$, which is known as the Stokes shift or Raman shift. The amount of frequency shift is dependent on the material, and it has a broad bandwidth gain of more than 40~THz with a Stokes shift of 13~THz~\cite{agrawal2007nonlinear} in silica glass.  The use of SRS in microcavities is attractive for such applications as Raman lasers~\cite{Spillane2002,Min:03,Kippenberg:04,1366398,Grudinin:07,Grudinin:08efficient,doi:10.1063/1.4820133,Latawiec:15,PhysRevLett.105.143903,Lin:16phase}, sensing~\cite{Li14102014} and self-frequency shift devices~\cite{PhysRevLett.116.103902,Yi:16}. The first observation of SRS in a silica microcavity was demonstrated in a microsphere~\cite{Spillane2002,Min:03}. Many theoretical and experimental studies were subsequently conducted in silica toroids~\cite{Kippenberg:04,1366398,doi:10.1063/1.4820133,doi:10.1063/1.2120921,Kato:17}, rods~\cite{doi:10.1063/1.4809781}, and bottle resonators~\cite{1882-0786-8-9-092001}. Recently, the influence of the SRS process on  microresonator Kerr comb generation has attracted a lot of attention as regards finding a way to enhance or suppress the SRS process. Studies have been undertaken on the influence of Raman scattering on soliton and Kerr comb generation~\cite{Tanabe:17,PhysRevA.92.033851,Yi:16}, the nonlinear coherent interaction between a Kerr comb~\cite{PhysRevA.92.043818,Lin:16phase} and a Stokes soliton~\cite{Yang2017}, and a transverse mode interaction via an SRS comb~\cite{Kato:17}.

In this work, we studied the competition between modulation instability (MI) and Raman gain theoretically and experimentally.  First we discuss the theory of gain competition to explain the transition between a Kerr comb and SRS and then calculate the system with the Lugiato-Lefever equation (LLE)~\cite{lugiato1987spatial}.  Next, we performed an experimental demonstration of the transition from the FWM dominant Kerr comb to the SRS dominant state in a silica toroid microcavity. Although some previous studies have reported on the transition between parametric oscillation and SRS~\cite{doi:10.1063/1.2120921,PhysRevLett.93.083904}, these studies only focused on a comparison of the maximum gains of the MI and SRS, and did not consider the resonance effect of the microcavity system. Therefore, the competition between a multi-FSR comb and a Raman state has yet to be studied in detail.

The paper is organized as follows.  Section~\ref{sec:2} presents a theoretical and analytical discussion of the competition between the FWM and SRS processes in a microcavity. Section~\ref{sec:3} describes numerical studies based on LLE. Section~\ref{sec:4} reports an experiment that was conducted to demonstrate the transition between the FWM dominant Kerr comb state and the SRS dominant Raman comb state. Section~\ref{sec:5} provides our conclusion.

\section{Theory and analysis}\label{sec:2}
\subsection{Theory of the MI and Raman gains in a silica microcavity}\label{sec:analysis}
MI is a phenomenon that induces a parametric oscillation from vacuum fluctuations in nonlinear materials. When the system is pumped with a continuous wave (CW) input, sidebands with frequencies other than the pump frequency are generated. When the phenomenon takes place in a microcavity system it initiates cascaded FWM and forms coherent broad spectral Kerr combs.  The gain spectrum $g(\Omega)$ of the MI including the loss in an optical fiber is derived from the nonlinear Schr\"{o}dinger equation (NLSE), as described in~\cite{agrawal2007nonlinear},
\begin{equation}
g_{\mathrm{fib}}(\Omega)=-\alpha_{\mathrm{fib}} + \left| \beta_2 \Omega \right|\sqrt{\Omega_c^2 - \Omega^2}, \label{Eq.1}
\end{equation}
where,
\begin{equation}
\Omega_c^2=\frac{4\gamma P_0}{\left| \beta_2 \right|}, \label{Eq.2}
\end{equation}
is the frequency of the gain peak. $\Omega$, $\beta_2$, $\gamma$, $\alpha_{\mathrm{fib}}$, and $P_0$ are the modulation frequency, second-order dispersion, nonlinear coefficient, propagation loss, and optical power, respectively.  On the other hand, MI gain in an optical microcavity is obtained from the LLE~~\cite{lugiato1987spatial}, which is an expansion of the NLSE used to describe the linear and nonlinear dynamics in an optical cavity.  The equation is given as~\cite{Torres-Company:14, Haelterman:92},

\begin{equation}
g_{\mathrm{cav}}(\Omega)=-\alpha_{\mathrm{cav}} + \sqrt{(\gamma L P_0)^2 -(\delta_{\mathrm{miss}})^2}, \label{Eq.3}
\end{equation}
where,
\begin{equation}
\delta_{\mathrm{miss}} = \delta_0-\frac{\beta_2}{2} L \Omega^2 -2\gamma L P_0, \label{Eq.4}
\end{equation}
is the phase-mismatch due to the detuning, dispersion and nonlinear phase shift. $\alpha_{\mathrm{cav}}$, $L$, and $\delta_0=t_R(\omega_0-\omega_p)$, are the loss of the cavity per roundtrip, the cavity length, and the phase detuning of the input frequency $\omega_p$ to the resonance frequency $\omega_0$ ($t_R$ is the cavity roundtrip time). In both cases, the higher order dispersions are neglected for simplicity.

\begin{figure}[b]
	\centering
	\includegraphics{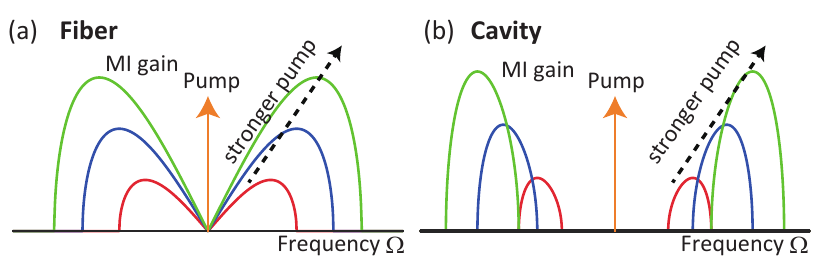}
	\caption{MI gain spectra for a fiber (a) and microcavity (b) system with different pump powers.  The peak of the MI gain shifts away from the pump frequency in both cases, but the gain is zero close to the pump frequency in the cavity case.}
	\label{fig:1}
\end{figure}
Figure~\ref{fig:1} shows these gain spectra for an optical fiber and cavity system.  One important difference between these two systems is that the MI gain is continuously present for the fiber system but it is absent at frequencies close to the pump frequency for the cavity system.  This is because of the presence of the cavity detuning, where the solution of Eq.~(\ref{Eq.3}) will be imaginary. The input power influences changes the phase-mismatch term, so the gap becomes larger when the input power is larger.

Another difference between these two systems is that the microcavity system has a discrete density-of-states; namely discrete longitudinal resonance frequencies separated by the FSR. This causes unique behavior with respect to the equidistantly spaced FWM generation that results from MI gain in a microcavity system. With a large cavity such as a fiber based ring cavity (i.e. a small FSR system), the MI gain always overlaps a number of longitudinal modes.  As a result, the frequency of the spectrum envelope of the generated FWM continuously shifts away from the pump when we increase the input power.  On the other hand, with a small microcavity system (i.e. a large FSR system), there is a large possibility that the MI gain will be located between the longitudinal modes.  As a result, none of the resonance frequencies of the cavity system receives the gain, and FWM generation is suppressed.  So now the question is; "How would this system behave if the cavity exhibited Raman gain at the same time?".

\begin{figure}[b]
	\centering
	\includegraphics{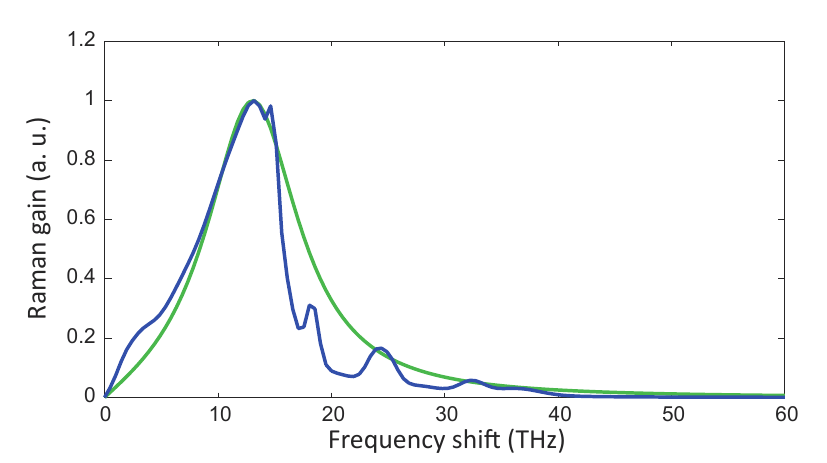}
	\caption{Raman gain spectrum in silica. Blue and green solid lines show an intermediate-broadening model and a single-damped-oscillator model, respectively~\cite{Hollenbeck:02}.}
	\label{fig:2}
\end{figure}

It is well known that silica has a broad Raman gain, and as a result, SRS may easily occur in a silica microcavity. The Raman gain $g_{\mathrm{R}}$ per roundtrip is given as~\cite{agrawal2007nonlinear},
\begin{equation}
g_{\mathrm{R}} = -\alpha + g_{\mathrm{bulk}}^{\mathrm{R}} \frac{P_0}{A_{\mathrm{eff}}} L_{\mathrm{eff}}, \label{Eq.5}
\end{equation}
\begin{equation}
L_{\mathrm{eff}} = \frac{1}{\alpha} [1-\exp(-\alpha L)], \label{Eq.6}
\end{equation}
where $g_{\mathrm{bulk}}^{\mathrm{R}}=0.6\times10^{-13}$ m/W~\cite{agrawal2007nonlinear} is the bulk Raman gain of silica at a pump wavelength of 1550~nm, and $A_{\mathrm{eff}}$ is the effective mode area. $L_{\mathrm{eff}}$ is the effective length determined by the propagation loss $\alpha$. Since the Raman gain spectrum is broad and its full-width at half-maximum (FWHM) covers more than 10~THz, as shown in Fig.~\ref{fig:2}, the Raman gain spectrum always covers multi-FSRs of the WGM microcavity system even when the cavity size is relatively small.
As a result of the broad Raman gain, the cavity always exhibits SRS gain when the system is CW pumped.

\begin{figure}[t]
	\centering
	\includegraphics{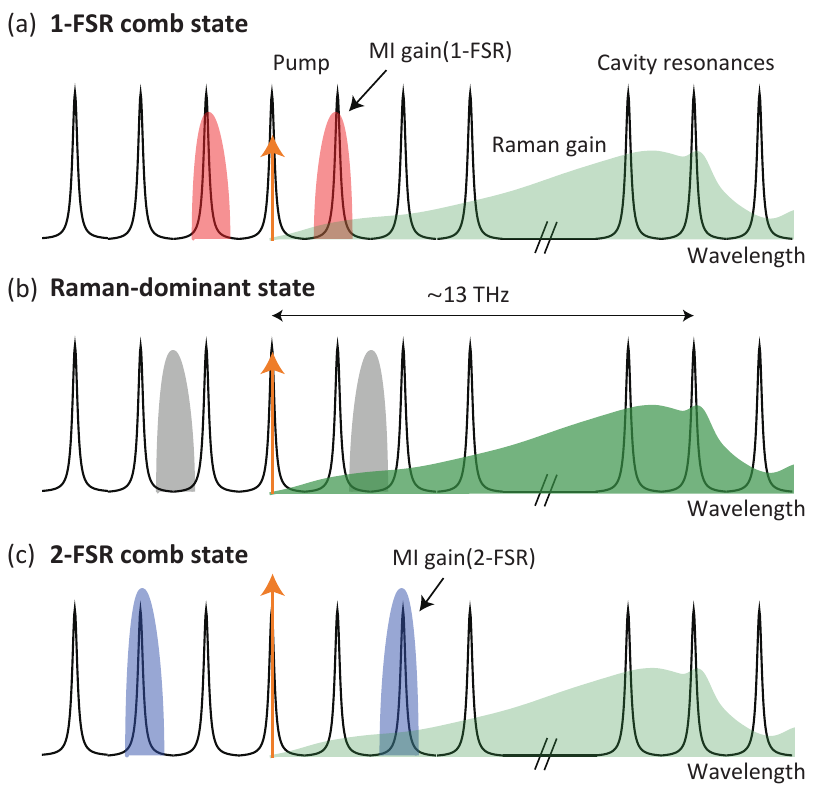}
	\caption{Schematic illustration explaining the competition between the MI and Raman gains in a small silica microcavity system.  (a) At a low pump power.  The peak of the MI gain overlaps a pair of longitudinal modes 1-FSR from the pump frequency.  (b) At a medium pump power.  The peak of the MI gain is between the longitudinal modes of the small microcavity.  The longitudinal modes overlaps the broad Raman gain.  (c) At a high pump power.  The peak of the MI gain overlaps a pair of longitudinal modes 2-FSR from the pump frequency.}
	\label{fig:3}
\end{figure}
Based on the above discussion, it is easy to understand that the gain competition between the FWM and SRS occurs in a relatively small WGM silica microcavity, as illustrated in Fig.~\ref{fig:3}.  When the input power is small, the MI gain overlaps a pair of resonances 1-FSR from the pump frequency.  Since the MI gain is higher than the Raman gain, FWM generation is dominant.  As a result, we can expect to obtain a 1-FSR comb state, where cascaded 1-FSR FWM occurs while the SRS process is suppressed [Fig.~\ref{fig:3}(a)].  When we increase the pump power, the MI gain shifts away from the pump frequency as shown in Fig.~\ref{fig:1}.  Then the MI gain may locate between adjacent resonances in the microcavity.  In this case, none of the microcavity resonance can receive the MI gain and so FWM generation is suppressed.  However, the SRS process can occur at a 13~THz red-shifted frequency due to the large bandwidth of the Raman gain.  As a result, FWM is suppressed and only the SRS process can occur, and the system exhibits an SRS dominant state [Fig.~\ref{fig:3}(b)].  When we further increase the pump power, the MI gain matches the next pair of resonances, namely 2-FSR from the pump.  Then those resonances receive MI gain and FWM is generated.  So the system should exhibit a 2-FSR comb state [Fig.~\ref{fig:3}(c)].

In the following subsection, we describe an analytical calculation that explains the phenomenon described above in more detail.

\subsection{Analysis of MI and Raman gains}\label{sec:analysis}
First, taking the optical bistable condition of a nonlinear cavity into account, we obtain the relationship between the input power $P_{\mathrm{in}}$ and the intracavity power $P_{\mathrm{0}}$ as,
\begin{equation}
\theta P_{\mathrm{in}} = (\gamma L)^2 P_0^3 -2\delta_0 \gamma L P_0^2 +(\delta_0^2 +\alpha^2)P_0, \label{Eq.7}
\end{equation}
where $\theta$ is the coupling coefficient between the cavity and the input waveguide.  Now we can obtain $P_0$ at a given $P_{\mathrm{in}}$ and $\delta_0$, and so we can calculate the MI and Raman gains as a function of the input power by using Eqs.~(\ref{Eq.3}) and (\ref{Eq.5}).

\begin{figure}[t]
	\centering
	\includegraphics{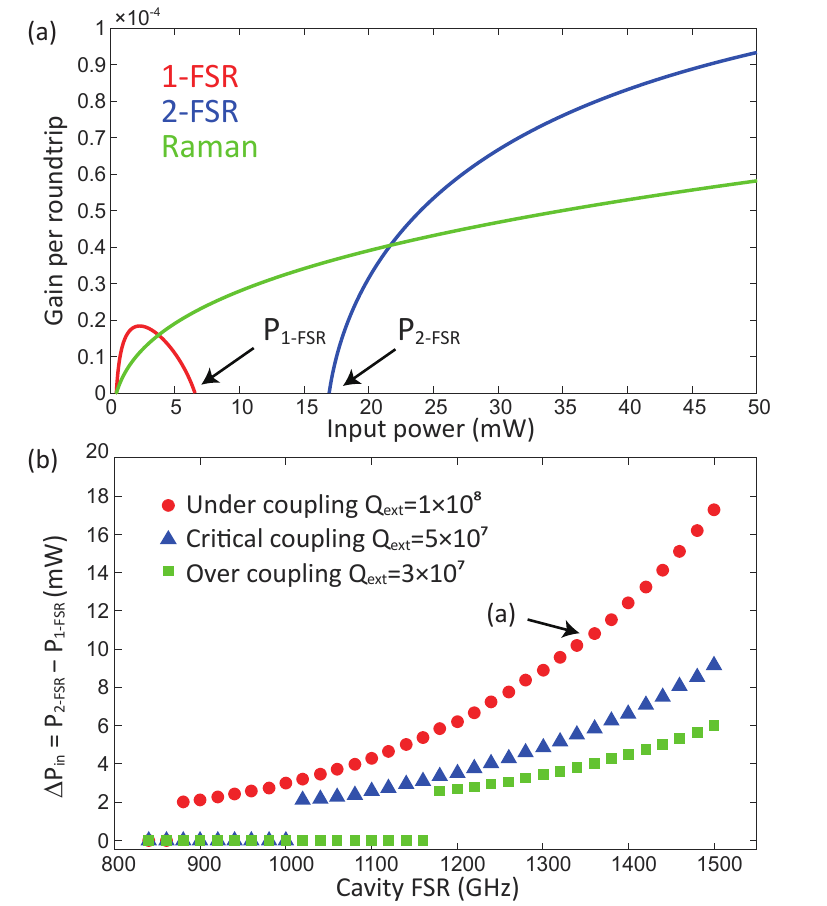}
	\caption{(a) Calculated MI gains at two different frequencies (1-FSR and 2-FSR from the pump), and the Raman gain per roundtrip as a function of the input power when $Q_\mathrm{ext}$ is $1\times10^8$.  Only the Raman gain interacts with the cavity modes when the input power is in the $6.5\sim17$~mW range, because the MI gain is located between the 1-FSR and 2-FSR resonant modes.  (b) $\Delta P_{\mathrm{in}}$ versus the cavity FSR at three different coupling conditions.  $\Delta P_\mathrm{in} = P_\mathrm{2\mathchar`-FSR}-P_\mathrm{1\mathchar`-FSR}$ is the allowed power range for obtaining the Raman-dominant state. The arrow is at the condition for (a).}
	\label{fig:4}
\end{figure}

Figure~\ref{fig:4}(a) shows the theoretical curves of MI gains at frequencies 1-FSR and 2-FSR from the pump frequency, along with the Raman gain.  We set the parameters as follows: pump wavelength $\lambda_p=1542$ nm, refractive index $n=1.44$, nonlinear refractive index $n_2=2.2\times10^{-20}$~$\mathrm{m^2/W}$, nonlinear coefficient $\gamma= 1.79\times10^{-2}$~$\mathrm{W^{-1} m^{-1}}$ ($\gamma$ is given as $n_2 \omega_0/c A_\mathrm{eff}$), intrinsic quality factor $Q_\mathrm{int}=5\times10^7$, external (coupling) quality factor $Q_\mathrm{ext}=1\times10^8$, and the phase detuning from cold cavity resonance $\delta_0=-5.4\times10^{-8}$.  It should be noted that an anomalous dispersion ($\beta_2<0$) is required for the scheme in Fig.~\ref{fig:1} to function (because phase-matching is satisfied only when $\beta_2<0$ under the condition $\delta_0 \sim 0$), so we assume a silica toroid microcavity, whose major and minor diameters are $\sim$ 50~$\mu$ m and $\sim$ 7~$\mu$ m, respectively.  (The fundamental mode of a small WGM microcavity usually exhibits a normal dispersion~\cite{Fujii:17})  The cavity FSR $\nu_\mathrm{FSR}(\Omega_\mathrm{FSR}/2\pi)=1350$~GHz, the second-order dispersion $\beta_2=-10$~$\mathrm{ps^2/km}$, the effective mode area $A_{\mathrm{eff}}=5$~$\mu$$\mathrm{m^2}$, all of which can be estimated by using the finite element method as the higher-order mode in a silica toroid microcavity.

Figure~\ref{fig:4} shows that there are three regions present, where 1-FSR MI gain, Raman gain, and 2-FSR MI gain are dominant with respect to the input power.  The cavity exhibits MI gain 1-FSR from the pump, but the gain at 1-FSR disappears as the input power increases.  Then a power regime apprears where only modes that overlap with the Raman gain receive the gain.  As we further increase the input power, the MI gain at a frequency 2-FSR from the pump becomes larger than the Raman gain.  So the result in Fig.~\ref{fig:4} directly supports our explanation in Fig.~\ref{fig:3}.

We define the maximum input power at which the gain at 1-FSR is equal to zero as $P_\mathrm{1\mathchar`-FSR}$, and the minimum input power at which the gain at 2-FSR is equal to zero as $P_\mathrm{2\mathchar`-FSR}$, as indicated in Fig.~\ref{fig:4}(a).  The difference between these two powers $\Delta P_\mathrm{in} = P_\mathrm{2\mathchar`-FSR}-P_\mathrm{1\mathchar`-FSR}$ is the range of the allowed input power for the system in a Raman-dominant state.  Figure~\ref{fig:4}(b) shows the $\Delta P_\mathrm{in}$ as a function of the cavity FSR for different coupling rates with the waveguide.  When $\Delta P_\mathrm{in}$ is zero, the system has no Raman-dominant region, but the MI gain at 2-FSR is dominant.  The calculation shows for example that the FSR of the cavity must be larger than 1000~GHz to obtain SRS when the system is operating in a critical coupling condition.  Hence, this analysis allows us obtain important information about the strategy for choosing the cavity diameter so that we obtain a Raman-dominant state without FWM.  The presence of the SRS comb will allow us to obtain broader bandwidth light~\cite{Jinnai:16,Fujii:17}.  The result shows that an under coupling condition ($Q_\mathrm{ext}>Q_\mathrm{int}$) in addition to the choise of a large-FSR (i.e. a small diameter microcavity) is suitable for obtaining a Raman-dominant region over a broad range.

We also investigated using a stronger pump, to determine whether this unique feature exists in the transition from 2-FSR to 3-FSR state.  However, the bandwidth of the MI gain is much broader at a higher pump power, and the MI gain always overlaps the longitudinal modes of the resonance, and it is not possible to find a Raman-dominant region with a realistic cavity diameter.

\section{Numerical calculation based on LLE}\label{sec:3}
As mentioned in the previous section, the LLE model well describes the dynamics of a nonlinear microcavity and should facilitate a more accurate the discussion of MI.  We consider the Raman effect on LLE as follows~\cite{PhysRevA.92.043818,Lin:16,Matsko:11},
\begin{equation}
\begin{split}
\frac{\partial E(\phi,t)}{\partial t} = -\left( \frac{\kappa_\mathrm{tot}}{2} +i \delta_0 \nu_\mathrm{FSR} \right)E +i v_g \sum_{k=2}^{\infty}(i \Omega_\mathrm{FSR})^k \frac{\beta_k}{k!}\frac{\partial^k E}{\partial \phi^k}  \\
+ i v_g \gamma f_{\mathrm{R}} \left[ E(\phi,t) \int h_{\mathrm{R}}(\phi'/\Omega_\mathrm{FSR})|E(\phi-\phi',t)|^2 d\phi' \right]\\
+ i v_g \gamma (1-f_{\mathrm{R}}) |E|^2 E 
+ \sqrt{\kappa_\mathrm{ext} \nu_\mathrm{FSR} P_{\mathrm{in}}} , \label{Eq.8}
\end{split}
\end{equation}
where $\phi $ is the azimuthal angle along the circumference of the cavity, $t$ is the time that describes the evolution of the field envelope, $\kappa_\mathrm{tot}=\kappa_\mathrm{int}+\kappa_\mathrm{ext}$ is the total decay rate given by sum of the intrinsic loss and external coupling rate, and $v_g$ is the group velocity. It should be noted that $E(\phi,t)$ are the slowly-varying fields and $|E|^2$ are normalized to the optical power. The second term of the right hand side describes the dispersion at all orders although higher order dispersions (i.e. 3rd, 4th) are neglected in the following calculation. The third and fourth terms on the right hand side describe the Raman and Kerr effects, respectively.   $f_{\mathrm{R}}$ is the fractional contribution of the delayed Raman response, which is known as $f_{\mathrm{R}}=0.18$ in silica, and $h_{\mathrm{R}}$ is the Raman response function, which is given as,
\begin{equation}\label{Eq.9}
h_{\mathrm{R}}(t)=\frac{\tau_1^2+\tau_2^2}{\tau_1 \tau_2^2} \exp{\left(-\frac{t}{\tau_2}\right)} \sin{\left(\frac{t}{\tau_1}\right)},
\end{equation}
where $\tau_1=12.2$~fs, and $\tau_2=32$~fs~\cite{agrawal2007nonlinear}. The Fourier transformed gain function is shown by a green line in Fig.~\ref{fig:2}.
\begin{figure}[!ht]
	\centering
	\includegraphics{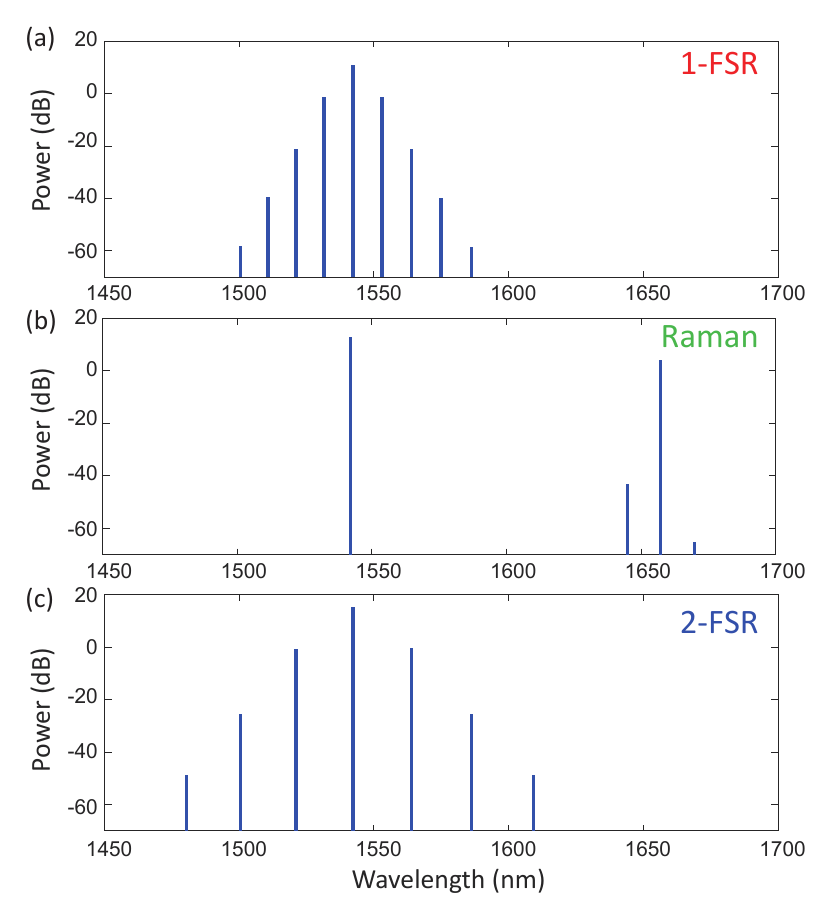}
	\caption{Output spectrum of a silica WGM microcavity obtained with LLE for different input powers. (a) Output spectrum when the system is pumped with a 2.5~mW input.  1-FSR Turing pattern comb is observed.  (b) At 10~mW input.  The peak of the output spectrum at 1657~nm wavelength corresponds to the peak of the Raman shift of silica. (c) At 40~mW input. A 2-FSR Turing pattern comb is obtained.}
	\label{fig:5}
\end{figure}

First, we set the input power at 2.5~mW to obtain the maximum MI gain at a 1-FSR frequency spacing.  The other parameters are the same as those we used in the analysis in section~\ref{sec:2}.\ref{sec:analysis}.  The calculated spectrum is shown in Fig.~\ref{fig:5}(a) and it is a stable comb spectrum at a 1-FSR spacing as we expected.  Specifically, it is a Turing pattern comb, which is the result of cascaded FWM process at a spacing of 1-FSR. Please note that we do not observe any spectrum component that is a result of the Raman process.  This shows that the FWM comb is dominant in this state.  Next, we increase the pump power to 10~mW and observe the output spectrum [Fig.~\ref{fig:5}(b)].  The result confirms that only SRS is generated at frequency $\sim$13~THz red-shifted from the pump.  It shows that the Raman gain outperforms the MI gain, and the SRS process is taking place, which agrees well with the situation discussed in Fig.~\ref{fig:3}(b). Finally, we obtain a stable 2-FSR Turing pattern  comb at an input power of 40 mW~[Fig.~\ref{fig:5}(c)], where no SRS process is observed. The observation agrees perfectly with our prediction and the analytical calculation.

We would like to note that these transitions can also be observed by changing different parameters such as the coupling $Q$ or the detuning of the input laser light.  This result offers the possibility of switching between the Kerr comb and the SRS comb using the same cavity simply by changing the input power or the coupling $Q$.

\section{Experiment}\label{sec:4}
\begin{figure}[!t]
	\centering
	\includegraphics{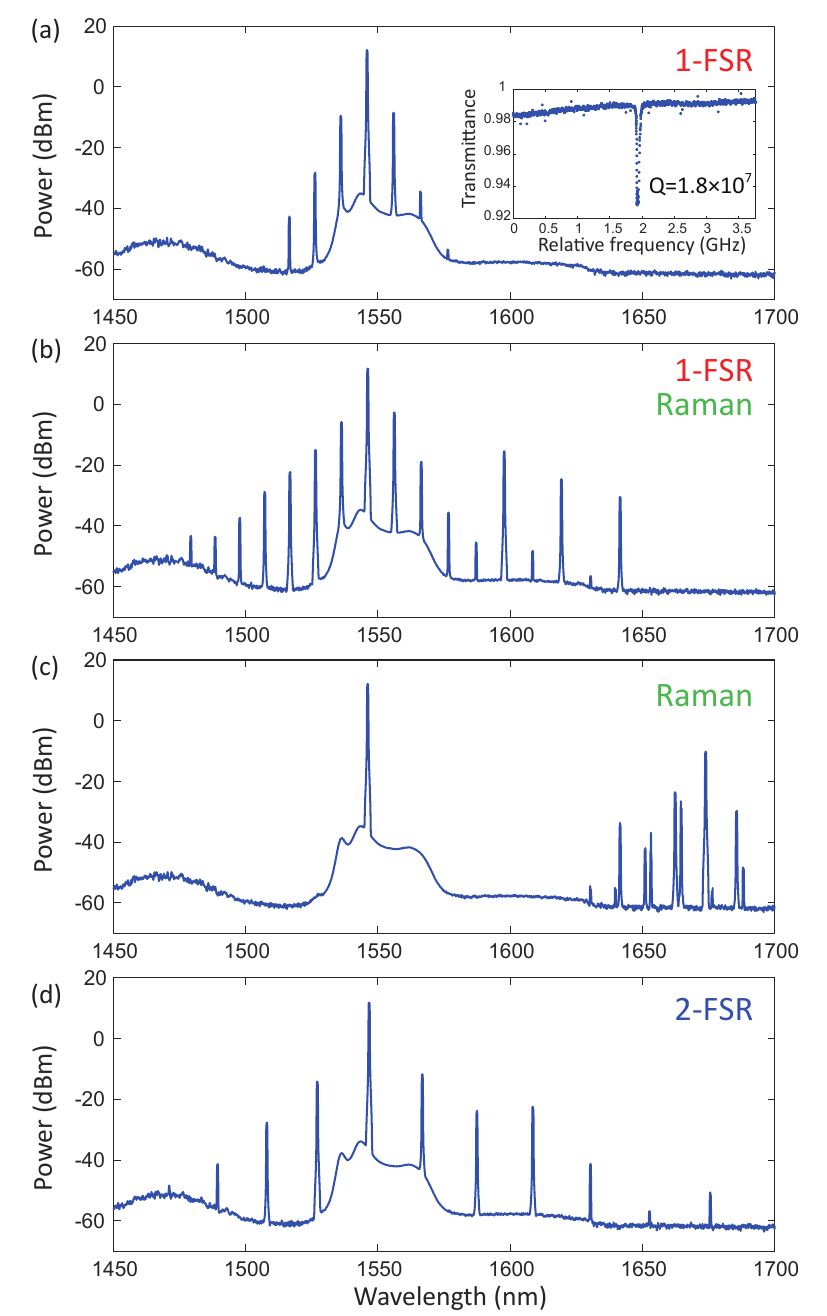}
	\caption{Measured output spectra from a silica toroid microcavity pumped at different detuning (i.e. different coupling power).  The wavelength of the input laser is swept from a shorter (a) to a longer (d) wavelength.  (a) When the pump is largely blue detuned, a 1-FSR comb is obtained. The inset shows the measured transmission spectrum.  (b) When the wavelength of the input laser is lengthened (i.e. smaller detuning), the spectral envelope of the 1-FSR comb becomes broader.  In addition, long wavelength components start to appear.  (c) When the detuning is smaller than (b), the FWM lines disappear and only the SRS comb is observed. (d) When the detuning is at its smallest, a 2-FSR comb is observed.}
	\label{fig:6}
\end{figure}

Finally, we describe the experiment we performed to confirm the above analysis. We fabricated a silica toroid microcavity with a major diameter of $\sim$50~$\mu$ m and a minor diameter of $\sim$7~$\mu$ m.  The measured $Q$ factor of the pump mode was $1.8\times10^7$ as shown in the inset of Fig.~\ref{fig:6}(a). The pumped resonance exhibits a slight mode splitting due to backscattering of light~\cite{Fujii:17cwccw}. We coupled the CW input to the microcavity with a tapered optical fiber setup, where the transmitted optical power and the output spectrum were recorded with a power meter and an optical spectrum analyzer. In the experiment, we changed the laser detuning over the resonance, which is equivalent to changing the cavity coupling power. The input pump power was 250~mW and the operating laser wavelength was $\sim$1546~nm.

The experimental results are shown in Fig.~\ref{fig:6}.  When we gradually changed the input laser detuning from a short to a long wavelength, we first obtained a stable output spectrum as shown in Fig.~\ref{fig:6}(a). Although we did not monitor the laser detuning, the experiment is performed in effectively blue detuned region. Therefore it is thermally stable~\cite{Carmon:04}. It exhibited a 1-FSR comb, which corresponds to the case in Fig.~\ref{fig:5}(a).  It is a Turing pattern comb since the detuning of the input light is located on the effectively blue detuned side of the cavity resonance. To increase the coupling power, we then slightly changed the detuning of the pump to a longer wavelength.  Then the spectrum changes, and a broad spectrum is obtained as shown in Fig.~\ref{fig:6}(b).  The spectrum is broader than that in Fig.~\ref{fig:6}(a), because the cavity is pumped with a stronger laser field as a result of stronger coupling of the input light due to smaller detuning from the cavity resonance.  It also shows some evidence of the Raman process, since the spectrum is broadened towards the longer wavelength side.  In addition, it shows evidence of the transition from a 1-FSR comb to an SRS comb.  When we further changed the detuning of the input, the spectrum exhibited a great change where the Kerr comb disappeared and an SRS comb appeared, which shows that the system is now in a Raman-dominant state [Fig.~\ref{fig:6}(c)].  We observe non-equidistant modes, which is the result of the coupling with different transverse modes that could occur during the SRS process~\cite{Kato:17}.  When we detune the input further, which corresponds to the highest coupled power in Fig.~\ref{fig:6}, SRS comb disappeared and an 2-FSR comb was observed as shown in Fig.~\ref{fig:6}(d). It should be noted that the wavelength component around 1675~nm is due to the remaining Raman effect, though the power is very weak. Although we often observe the coexistence of the FWM and SRS processes~\cite{Jinnai:16,Fujii:17}, particularly when we carefully design the dispersion of the cavity system, here we clearly observed the transition from FWM to the Raman state and then back to the FWM state.  The experimental results agree well with the analysis and the simulation, which indicates that the transition from FWM to the Raman-dominant state is present in a silica microcavity system when we try to switch between different-FSR Kerr comb states.

\section{Conclusion}\label{sec:5}
We theoretically and experimentally demonstrated the gain transition between FWM and SRS dominant states in a silica WGM microcavity. Steady-state analysis and simulation using LLE allowed us to reveal the phenomenon, where a Raman-dominant state is present between Kerr comb states with different-FSRs due to the broadband Raman gain of silica.  Although a Raman comb is of high interest in terms of extending the wavelength regime of the microresonator frequency combs, this study revealed that we need to choose the cavity size, the gap between the cavity and the waveguide, and the input power carefully, in order to obtain a Raman-dominant state.  This finding will also help studies on efficient Raman lasing or sensing applications in silica microcavities.

Just before submitting this paper, we came across a paper studying the competition between the Raman and Kerr effects in crystalline microresonator~\cite{YoshitomoComp}.  It shows that the SRS comb can be suppressed when the Raman gain is narrow, by designing the FSR of the cavity system.  On the other hand, our study deals with finding the condition to obtain a broadband SRS comb in a silica microcavity with a large FSR spacing.

\section*{Funding}
This work was supported by KAKENHI (\#15H05429) and the Photon Frontier Network Program, both established by the Ministry of Education, Culture, Sports, Science and Technology (MEXT).

\section*{Acknowledgment}
We thank T. Kobatake and A.-C Jinnai for helpful discussions.

%\bibliography{bibfile_fwm_srs_fujii}

\end{document}